\documentclass[%
 reprint,
superscriptaddress,
nofootinbib,
 amsmath,amssymb,
 aps,
 prl,
]{revtex4-2}

\usepackage{xcolor}
\usepackage{graphicx}%
\usepackage{dcolumn}%
\usepackage{bm}%
\usepackage{times}%
\usepackage{hyperref}%
\usepackage[normalem]{ulem}
\hypersetup{
  colorlinks=true,        %
  linkcolor=blue,         %
  citecolor=cyan,         %
  urlcolor=cyan            %
}

\begin{document}

\title{Reconnection-powered fast radio transients from coalescing neutron star
binaries}

\author{Elias R. Most}
\email{emost@princeton.edu}
\affiliation{Princeton Center for Theoretical Science, Princeton University, Princeton, NJ 08544, USA}
\affiliation{Princeton Gravity Initiative, Princeton University, Princeton, NJ 08544, USA}
\affiliation{School of Natural Sciences, Institute for Advanced Study, Princeton, NJ 08540, USA}

\author{Alexander A. Philippov}
\affiliation{Department of Physics, University of Maryland, College Park, MD 20742, USA}

\date{\today}
\begin{abstract}
It is an open question whether and how gravitational wave events involving
neutron stars can be preceded by electromagnetic counterparts. This work
shows that the collision of two neutron stars with magnetic fields well
below magnetar-level strengths can produce millisecond
Fast-Radio-Burst-like transients.  Using global force-free electrodynamics
simulations, we identify the coherent emission mechanism that might operate
in the common magnetosphere of a binary neutron star system prior to
merger. We predict that the emission show have frequencies in the range of
$10-20\,\rm GHz$ for magnetic fields of $B^{\ast}=10^{11}\,\rm G$ at the
surfaces of the stars.

\end{abstract}

\maketitle

\textit{Introduction.}
Gravitational wave events involving the collision of two neutron stars \cite{LIGOScientific:2017vwq,LIGOScientific:2020aai},
are a rich playground for multi-messenger astronomy \cite{LIGOScientific:2017ync,Kasliwal:2017ngb}.
They a broad range of transients in the optical \& infrared \cite{Cowperthwaite:2017dyu,Chornock:2017sdf,Villar:2017wcc,Nicholl:2017ahq,Troja:2018ruz,Tanvir:2017pws,Drout:2017ijr},
gamma- \cite{LIGOScientific:2017zic,Savchenko:2017ffs} and X-ray \cite{Troja:2017nqp,Margutti:2017cjl,Margutti:2018xqd,Hajela:2019mjy}, as well as radio \cite{Hallinan:2017woc,Alexander:2017aly,Ghirlanda:2018uyx,Mooley:2017enz,Mooley:2018qfh} bands.
While at present, it has only been possible to detect gravitational waves
emitted during the inspiral of two neutron stars, the above electromagnetic transients are thought to be sourced during the merger and post-merger phase,  respectively during and after the collision (see, e.g.,  \cite{Metzger:2019zeh,Radice:2020ddv} for recent reviews).
On the other hand, the presence of potentially strong magnetic fields in neutron stars has led several studies to propose that there might be yet another electromagnetic counterpart, that -- different from the others -- would be sourced prior to merger. Such an event is commonly referred to as an \textit{electromagnetic precursor}. 
In fact, recent (indirect) observational evidence has pointed towards the existence of such precursor events \cite{Xiao:2022quv}, which additionally claimed the presence of magnetar-level field strengths in the inspiraling system.
Typical models generally have predicted emission across the entire electromagnetic spectrum involving gamma-ray \cite{Tsang:2011ad,Metzger:2016mqu,Zhang:2022qtd}, radio \cite{Hansen:2000am,Wang:2016dgs,Lyutikov:2018nti,Sridhar:2020uez} and/or X-ray transients \cite{Beloborodov:2020ylo}. 
While these models have mostly been analytical, a number of studies have attempted to numerically model various global scenarios producing these transients. They include emission associated with the motion of the neutron stars \cite{Carrasco:2020sxg}, flares in the magnetosphere \cite{Most:2020ami,Most:2022ojl} [see also \cite{Beloborodov:2020ylo,Cherkis:2021vto}], dissipation in the common magnetosphere of the two stars \cite{Palenzuela:2013hu,Palenzuela:2013kra,Ponce:2014hha}, or balding of the black hole formed in a prompt collapse scenario \cite{Nathanail:2020fkp} [see also \cite{Lehner:2011aa,Palenzuela:2012my,Dionysopoulou:2012zv,Most:2018abt,Bransgrove:2021heo}]. Similar numerical studies have also been performed for black hole -- neutron star systems \cite{Paschalidis:2013jsa,East:2021spd,Carrasco:2021jja}. \\
Different from all previous studies -- analytical and numerical -- here we utilize the three-dimensional dynamics modeled in global simulations of the common magnetosphere to clearly establish which coherent emission mechanism is likely operating, and under what conditions. We demonstrate, for the first time, that electromagnetic flares emitted prior to merger can interact directly with the current sheets present in the binary's magnetosphere, causing large scale reconnection in the compressed orbital current sheet. Leveraging state-of-the-art models for linearly polarized Fast Radio Bursts (FRBs) \cite{Lyubarsky:2020qbp,Mahlmann:2022nnz}, we robustly link the interaction dynamics inside the common binary magnetosphere to an FRB-like millisecond transient at frequencies of $10-20\,\rm GHz$.\\

\begin{figure*}
    \centering
    \includegraphics[width=\textwidth]{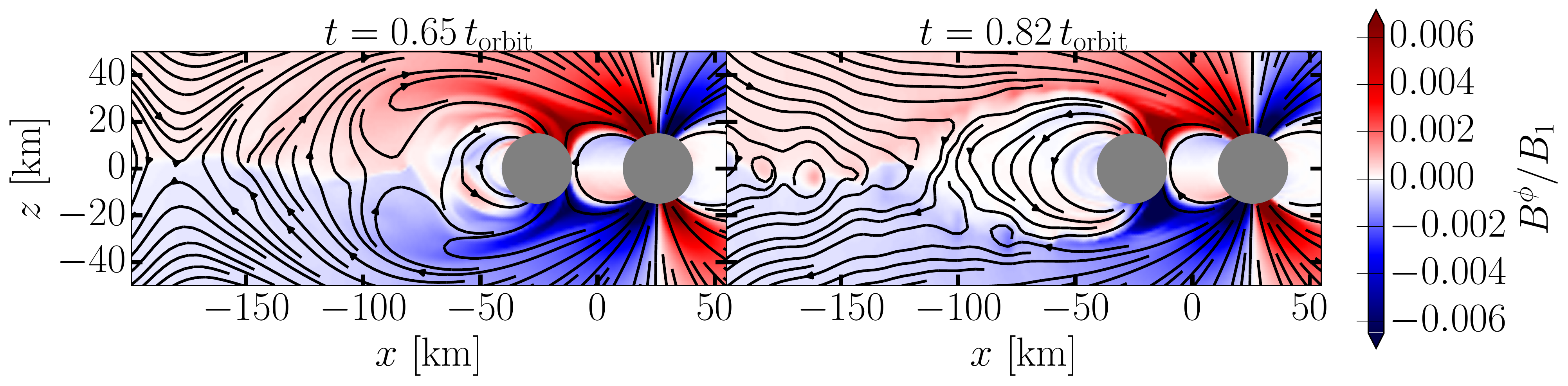}
    \caption{Representative interaction of electromagnetic precursor flares with the orbital current sheet (at $z=0\,\rm km$) for counter-aligned dipoles. \textit{(Left)} { Launching of the flares from overtwisted flux tubes. }\textit{(Right)} Reconnection in the current sheet and the formation of plasmoids. Shown in color is the out-of-plane component, $B^\phi$, of the magnetic field, which indicates the twist in the magnetosphere. The times are shown relative to the orbital period, $t_{\rm orbit}$, and field strength values are stated relative to the value $B_1$ at the surface of the right star.}
    \label{fig:Fig0}
\end{figure*}
~\\
\textit{Results.} In the following, we describe a novel mechanism capable of producing fast radio transients prior to the merger of two neutron stars. This reconnection-driven mechanism originates from the interaction of electromagnetic flares with the common binary neutron star magnetosphere.
We proceed in three steps. First, we utilize global force-free electrodynamics simulations to study a binary neutron star system featuring spinning stars with misaligned magnetic moments. Such an approach is appropriate to capture the dynamics of the highly conducting pair-plasma-filled binary magnetosphere \cite{Palenzuela:2010nf,Parfrey:2013gza,Carrasco:2019aas} (see the Appendix for details). Second, based on the
  simulation outcome we identify the likely candidate for a coherent radio
  emission mechanism. Finally, we use insights gained from previous kinetic
  simulations and analytic theory associated with the mechanism to predict the emission consistent with
condition probed in our global simulations. \\
We begin by illustrating the flaring mechanism and its interaction with the orbital current sheet. 
Neutron star binaries with oppositely oriented magnetic moments can feature connected magnetic flux tubes \cite{Piro:2012rq,Lai:2012qe}. Relative motion between the two stars, i.e. orbital motion or stellar spin, can lead to a twisting of those flux tubes. Once overtwisted, the connected field lines will snap and release a strong magnetic bubble \cite{Most:2020ami}. In an earlier work \cite{Most:2022ojl}, we have shown that this mechanism is relatively insensitive to the precise field geometry, strength or alignment, as long as the overall topology (magnetic moments pointing in opposite hemispheres) is preserved. Generically these flares can feature Poynting fluxes of $> 10^{42}\, \rm erg/s$ for field strengths $B\sim 10^{11}\,\rm G$. While the mechanism can in principle operate earlier, the flaring strength will only become observationally relevant in the final second before the collision \cite{Most:2020ami,Most:2022ojl}.
Crucially, the missing component in these studies has been the identification of a self-consistent emission mechanism applicable to this system. To alleviate this, it is necessary to track the subsequent evolution of the flare after it has been launched from the two stars. For ease of demonstration, we first focus on an idealized system consisting of two perfectly anti-aligned magnetic dipoles, with one of them being $50\times$ less strong than the other. In the second step, we will focus on a more realistic system.
Starting with the left panel of Fig. \ref{fig:Fig0}, we can see that the relative orbital motion induces a twist of the connected flux tubes, leading to the emission of two flares. Because of the unequal field strengths of the dipoles, the flares are strongly beamed towards the orbital plane \cite{Most:2022ojl}. They will ultimately hit the orbital current sheet and enhance reconnection (right panel). We can see that the latter leads to the appearance of plasmoids (magnetic islands), the merger of which is associated with the production of low-frequency radiation \cite{Lyubarsky:2018vrk,Philippov:2019qud}. While the properties of the emission will strongly depend on the ratio of the magnetic field strength in the flare to the background field in the current sheet, a similar mechanism has been proposed to explain linearly polarized FRBs from active magnetars \cite{Lyubarsky:2020qbp}. In the following, we will demonstrate that the same scenario naturally arises in generic binary configurations. In the third step, we will use those to infer the properties of the resulting emission.\\
\begin{figure*}
    \centering
    \includegraphics[width=\textwidth]{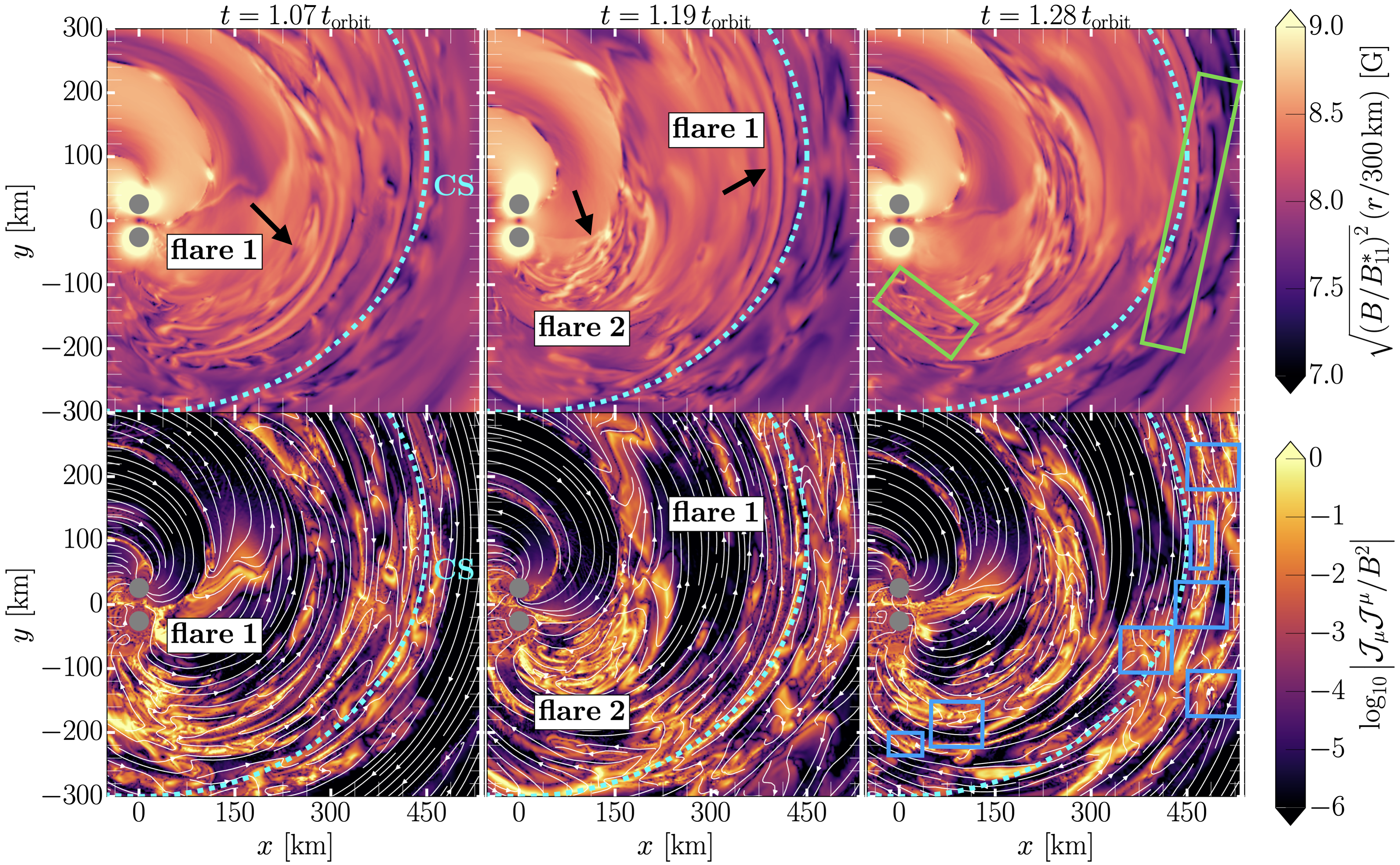}
    \caption{Interaction of electromagnetic precursor flares with the orbital current sheet (CS) in the orbital plane. A sequence of two flaring events is shown. As the flares propagate outwards they interact with the orbital current sheet. For clarity, only one stripe has been marked using a dashed cyan line. The top row shows the total magnetic field strength, $\sqrt{B^2}$, re-scaled by the cylindrical radius, $r$,
    and magnetic field strength, $B_{11}^\ast$, (in units of $10^{11}\,\rm G$) at the surface of the stars.
    The bottom row shows the magnitude of the electric four-current,
    $\mathcal{J}^\mu$, highlighting the current sheet. The interaction
    process triggers reconnection that leads to the formation of plasmoids
    (magnetic islands), some of which are highlighted by a green box.
    Flare-induced reconnection x-points are highlighted by blue
    boxes. All times, $t$, are stated relative to the orbital period, $t_{\rm orbit}$.}
    \label{fig:Fig1}
\end{figure*}
We now focus on a more realistic system, where one of the neutron stars is
rapidly counter-spinning and has an inclined magnetic field. Although most
neutron star binaries are thought to be nearly irrotational
\cite{Bildsten:1992my}, we specifically pick this configuration as it
leads to a sub-millisecond periodicity, potentially relevant for
FRB20201020A-type events \cite{Pastor-Marazuela:2022pnp}. More
specifically, in Fig. \ref{fig:Fig1} we consider the evolution in the
orbital plane showing both the magnetic field strength $\sqrt{B^2}$ and the
squared current density $\mathcal{J}_\mu \mathcal{J}^\mu$. Since the
flaring happens periodically with a combination of the orbital and spin
period of the stars \cite{Cherkis:2021vto}, this will highlight the
propagation of the flares. During the time interval shown, two flares will
be emitted, although we will largely focus on the first flare only. For
clarity, we have labeled them as \textit{flare 1/2}. Tracking their
evolution in time, we can identify three different phases: pre-interaction
(left panel), interaction (middle panel) and post-interaction phase (right
panel) with the highlighted part of the orbital current sheet.
Concentrating at the magnetic field strength, we can follow the propagation
of flare 1. After its inception it expands and approaches the orbital
current sheet. We point out that since the simulations have been performed
in the corotating frame, the flares have a counter-clockwise outwards
spiralling motion in this depiction.  Once the flare reaches the
highlighted part of the current sheet, we can see that the dissipation is
locally enhanced. This is most easily visible in the bottom row, which
indicates the (complex) current sheet structures.  We can see that the
interaction is associated with a compression of the current sheet. Finally,
after the flare has propagated past the sheet, small magnetic islands
(plasmoids) begin to appear (right column). They can be seen as local
overdensities (bright spots) both in the magnetic field
strength,$\sqrt{B^2}$, and in the co-moving current density,
$\sqrt{\mathcal{J}_\mu \mathcal{J}^\mu}$. In fact, due to the periodic
emission of flares, multiple interactions of flares with different parts of
the orbital current sheet will locally produce plasmoids throughout the
domain, as can be seen in the little bright spots present at all three
times shown in Fig. \ref{fig:Fig1}.  Some of those have already begun to
merge, although their physical size and full dynamics cannot be captured
with the force-free approach adopted in this work. Nonetheless, our
simulations clearly establish the viability of the current sheet-flare
interaction mechanism outlined above. In fact, the authors are not aware of
a global simulation demonstrating this mechanism even for an isolated star
(see \cite{Mahlmann:2022nnz} for a local simulation of the interaction
itself). In order to better demonstrate the flare-current sheet
interaction,
Fig. \ref{fig:Fig2} shows a zoom-in rendering of a flare hitting the
orbital current sheet. Both the toroidal field structure of the flare and the
deformation of the current sheet are clearly visible. Subsequently, the flare drags along the
orbital current sheet, compressing it and causing it to reconnect on large
(angular) scales. 
This implies that the emission caused by the merger of plasmoids will
likely cover a large spatial volume, enhancing its detection prospects
greatly \cite{Callister:2019biq}. \\

\textit{Emission process.} 
In the following, we want to characterize the emission we expect as a
result of the current sheet-flare interaction. This entails estimating
luminosity, duration and frequency of the event.  To do so, we will combine
global numerical findings, i.e., magnetic field strengths, spatial flare
extent and duration, with analytical expression for the small-scale plasma
physics processes not captured by the simulations
\cite{Lyubarsky:2020qbp,Mahlmann:2022nnz}. This way we are able to
   model unresolved reconnection microphysics not captured by artificial
 dissipation in our force-free approach.   We start out by describing the
luminosity and event duration. Fig. \ref{fig:Fig3} shows the Poynting flux
luminosity associated with the flares, $\mathcal{L}_{\rm EM}$, and the
dissipative luminosity, $\mathcal{L}_{\rm diss}$, associated with
reconnection (see \cite{Most:2022ojl} for details on how to compute these
quantities).  We can see that periodic flaring is present, with about four
flares being emitted per orbit. The exact periodicity can be computed based
on topological arguments \cite{Cherkis:2021vto}. We point out that this
rapid periodicity is a result of having counterspinning stars in the
binary, making it suitable to produce observable sub-ms variability.
Overall we find that $\mathcal L_{\rm EM} \simeq 10^{42}
\left(B^\ast_{11}\right)^2 \, \rm erg/s $ and $\mathcal L_{\rm diss} \simeq
-3\times 10^{41} \left(B^\ast_{11}\right)^2 \, \rm erg/s$, where
$B^\ast_{11}$ is the value of the strength of the field at the stellar
surface in units of $10^{11}\, \rm G$. We point out that the physical
  Lundquist number associated with this dissipation will be high so that reconnection will happen in the plasmoid regime \cite{Loureiro:2007gv,2009PhPl...16k2102B}.
From Fig. \ref{fig:Fig2} we read off
a light-crossing time across a flaring electromagnetic bubble of $\tau
\simeq 0.3\, \rm ms$.  From Fig. \ref{fig:Fig1} we infer that the magnetic
field strength in the flare at the light cylinder, $B_{\rm flare} \simeq
8\times 10^8 B^\ast_{11} \rm G$, is about $20$-times stronger than the
field strength in the binary wind, $B_{\rm wind}$. The Lorentz factor of
the plasma in the flaring pulse can be estimated as
\cite{Lyubarsky:2020qbp}, $\Gamma = \sqrt{{B_{\rm flare}}/{B_{\rm
wind}}}/2 \approx 2.2$. When the flare arrives at the current sheet, it
triggers violent compression and reconnection. 
This is different from intrinsic tearing of the elongated sheet in the quiescent binary magnetosphere, 
  which alone is not energetic enough to produce observable transients
  (and at lower frequencies than in a compressed sheet).
The magnetic flux in the
pulse is comparable to the flux within the stripe of the binary wind,
$\phi_{\rm flare}/\phi_{\rm wind}\sim {B_{\rm flare}}/{B_{\rm wind}}
(c\tau/R_{\rm LC})\approx {\rm {few}}$, where $R_{\rm LC}$ is the light
cylinder of the binary. In this case, local kinetic simulations of
pulse-current sheet interactions \cite{Mahlmann:2022nnz} show that the
dissipated power is of order of $\mathcal{E}_{\rm diss}\sim \beta_{\rm rec}
L_{\rm EM} \tau$, where $\beta_{\rm rec} \approx 0.1$ is the dimensionless
rate of magnetic reconnection (e.g., \cite{Sironi:2014jfa,Guo:2015cua,Guo:2020ejn},
see also \cite{Du:2021vij,French:2022zfv}). For our
scenario, this results in $\mathcal{E}_{\rm diss} \approx 10^{38}
\left(B^\ast_{11}\right)^2 {\rm erg}$. The tearing-mode instability results
in the current sheets fragmentation into a sequence of magnetic islands,
and their collisions source low-frequency fast magnetosonic waves
\cite{Lyubarsky:2018vrk,Philippov:2019qud}. Simulations show that a fixed
fraction of the power dissipated in the reconnection, $f\approx 2 \times
10^{-3}$, gets converted into the low-frequency radiation
\cite{Philippov:2019qud,Mahlmann:2022nnz}. For the luminosity we then
obtain $\mathcal{L}_{\rm radio} \approx 0.1 f L_{\rm EM} \approx 2\times
10^{38}\,\left(B^\ast_{11}\right)^2\, \rm erg/s$.  The observed signal
  duration is of order of a light-crossing time across a flaring bubble \cite{Mahlmann:2022nnz}, $\sim \tau \simeq 0.3\, \rm ms$, and,  hence, can give rise to sub-millisecond transients \cite{Pastor-Marazuela:2022pnp}. 

\begin{figure}
    \centering
    \includegraphics[width=0.45\textwidth]{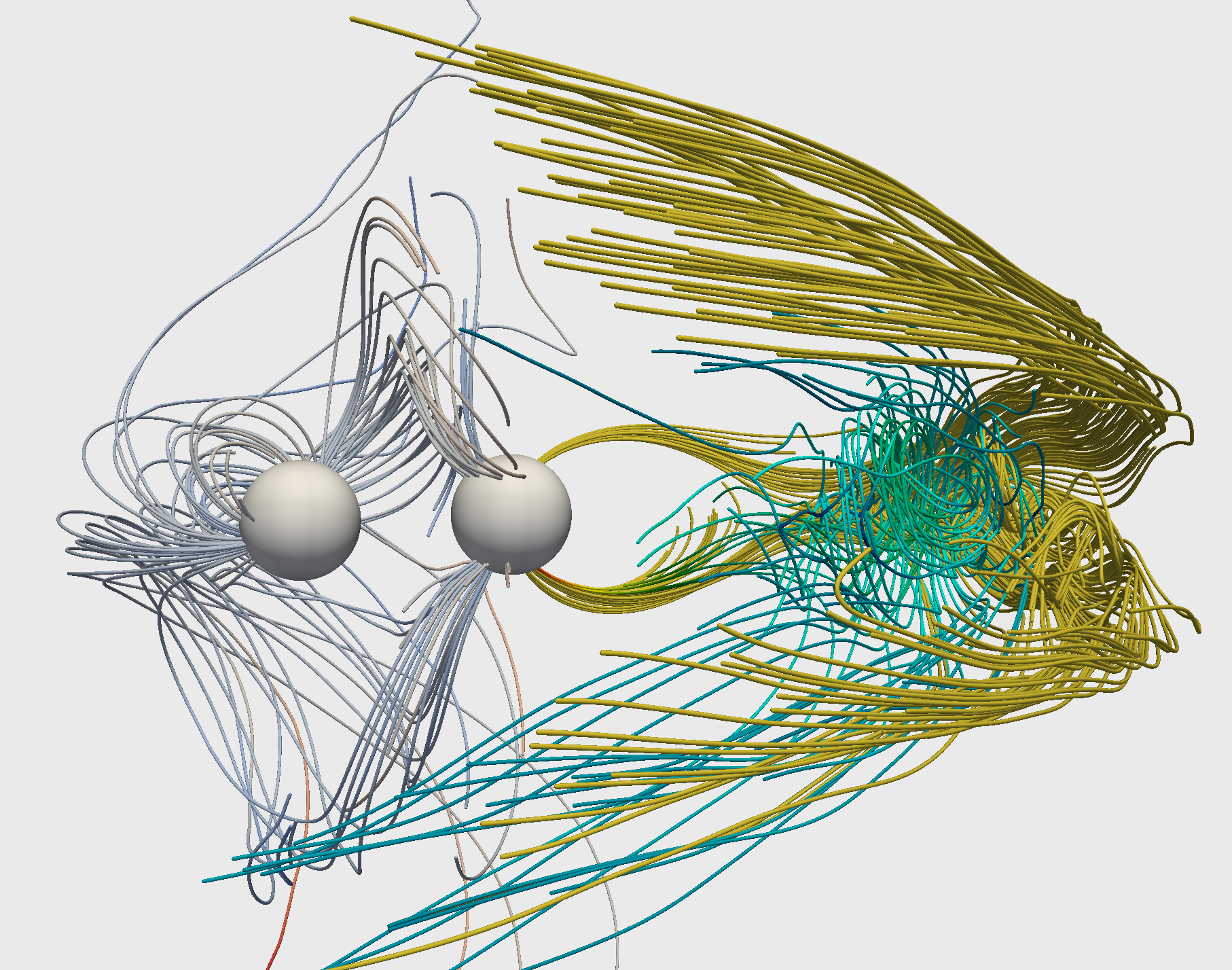}
    \caption{Zoom-in: selected magnetic field lines of the flare (blue) interacting with field lines of the orbital current sheet (yellow) at time $t \simeq 1.18 t_{\rm orbit}$. The field lines in the flare are predominantly toroidal, which begins to get compressed during the impact. For illustration
  purposes, only a subset of magnetic field lines is shown.}
    \label{fig:Fig2}
\end{figure}

In the next step, we need to establish that this reconnection-powered transient indeed occurs in the radio band. 
In the reference frame of the pulse, the characteristic frequency of the
emission associated with the merger of plasmoids, $\nu' = c/\left(2 \pi
\zeta d'_{\rm plasmoid}\right)$, is determined by the sizes of typical
plasmoids in the chain, $d'_{\rm plasmoid}=\xi r_L$, where $r_L=m_e c^2
\langle\gamma\rangle/(eB'_{\rm flare})$ is the Larmor radius of the
particles heated by reconnection up to averaged Lorentz factor
$\langle\gamma\rangle$, and $\zeta\approx 1, \xi\approx 10-100$ are
numerical factors. These length scales cannot be resolved in our global
force-free simulations, and would, in fact, require full kinetic
simulations. However, local small-scale simulations performed in the regime
of very fast synchrotron losses, applicable to the binary magnetosphere,
find a balance between particle energization by reconnection and
synchrotron cooling. This results in $\langle\gamma\rangle \approx
\gamma_{\rm rad}=\sqrt{3e \beta_{\rm rec}B'_{\rm flare}/(2 r^2_e B'^2_{\rm
flare}})$ (e.g., \cite{Mahlmann:2022nnz}). Following results of kinetic
simulations \cite{Mahlmann:2022nnz}, we adopt $\xi \zeta \approx 90$.
Transforming into the lab frame, this implies a frequency of the transient
\cite{Mahlmann:2022nnz,Lyubarsky:2020qbp},
\begin{align}
  \nu_{\rm FRB} =  \frac{1}{2 \pi \xi \zeta}
  \sqrt{\frac{2 r_e}{3 \beta_{\rm rec} c \Gamma} \omega_B^3} \simeq 16 \, \left(B^\ast_{11}\right)^{3/2}\,\rm GHz\,,
  \label{eqn:nu_flare}
\end{align}
where $\omega_B=eB_{\rm flare}/m_e c$ is the electron gyrofrequency in the
pulse frame, for the emission of fast radio transients from the late
inspiral of neutron star binaries. For varying parameters of the binary,
this frequency may decrease by a factor of a few, if the collision of the
pulse and current sheet happens closer to distance of $2 R_{\rm LC}$. As
the pulse propagates outwards and its magnetic field drops as $\sim 1/r$,
the emission frequency decreases, leading to the downward frequency drift
similar to a ``sad trambone'' effect observed in FRBs
\cite{Hessels:2018mvq,Josephy:2019ahz,Rajabi:2020uol}. This is
different from a ``trombone'' effect caused by orbital decay which will
push the flares to higher strengths of the magnetic field on collision with the orbital current sheet, causing the frequency to drift upwards between individual flaring events. We point out that this radio emission is distinct from X-ray transients produced in current
sheets behind the flaring bubble with higher field strengths
\cite{Beloborodov:2020ylo}. These would be produced earlier, but would
propagate together with the bubble escaping at relativistic speed and,
thus, arrive later to the observer with a delay of $\sim R_{\rm LC}/c\sim
1{\rm ms}.$ Using the dissipated power in Fig. \ref{fig:Fig3} as an upper
limit, we find X-ray luminosity is expected to be $L_{X}\lesssim 4\times
10^{41}\, \rm erg/s$. Lastly, we have also verified that fast magnetosonic
waves excited during the pulse-current sheet interaction can escape the
binary magnetosphere and wind as electromagnetic waves. An estimate of
these propagation effects is presented in the Appendix. %

\begin{figure}
    \centering
    \includegraphics[width=0.45\textwidth]{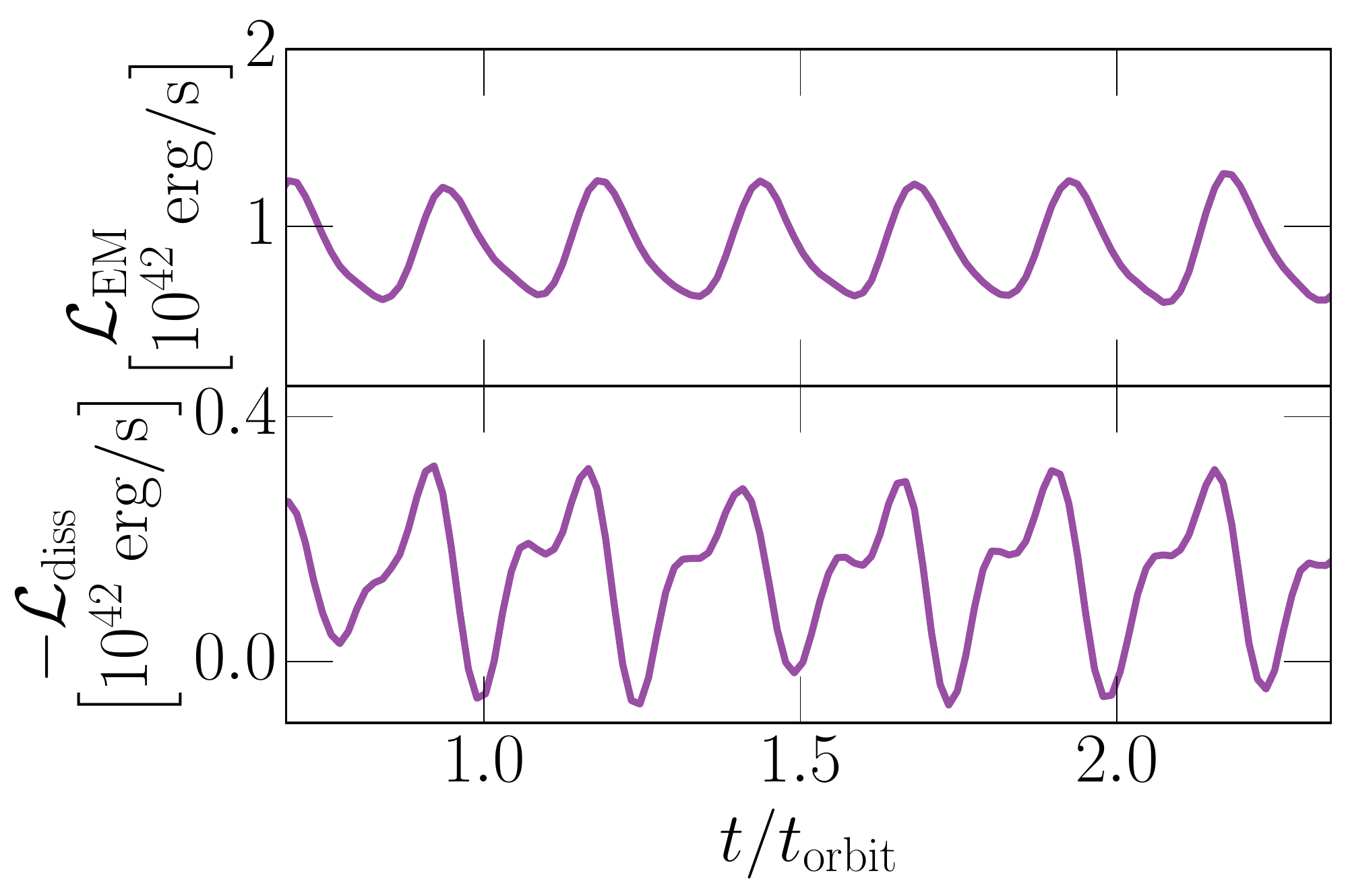}
    \caption{Electromagnetic power, $\mathcal{L}_{\rm EM}$, of the outgoing flare (top), and (approximately) dissipated power, $\mathcal{L}_{\rm diss}$, in the current sheet (bottom). These have been normalized for a $B^\ast \sim 10^{11}\, \rm G$ magnetic field at the surface of the stars.
    All times are stated relative to the orbital period, $t_{\rm orbit}$.}
    \label{fig:Fig3}
\end{figure}

\textit{Discussion.} ~ 
In this work, we have described a novel mechanism to produce FRB-like transients from the late inspiral of neutron star binaries. More specifically, we have shown that the presence of magnetic fields in excess of $10^{11}\, G$, will lead to the production of millisecond radio transients in the frequency range of $10-20\, \rm GHz$. This is based on identifying large scale interactions of electromagnetic flares with the orbital current sheet of the binary, akin to FRB production mechanisms for active magnetars \cite{Lyubarsky:2020qbp,Mahlmann:2022nnz}.
On the basis of our earlier findings \cite{Most:2020ami,Most:2022ojl}, we believe this mechanism to be generically applicable to many binary neutron star gravitational wave events provided that the magnetic field orientations of the stellar fields are loosely anti-aligned.
For such configurations, electromagnetic flares are being launched periodically from the system (see \cite{Cherkis:2021vto} for a detailed computation of the period), where flaring to first-order
will depend on the relative spin rate in the system \cite{Lai:2012qe}, which can be enhanced for counter-spinning neutron stars.
Hence, rapidly spinning binaries can produce sub-millisecond pulsations, similar to those observed in transients such as FRB20201020A \cite{Pastor-Marazuela:2022pnp}. Since most FRBs feature emission firmly $< 8\, \rm GHz$ (see, e.g., \cite{Petroff:2021wug} for a recent review), the mechanism proposed here would lead to an undiscovered sub-population of FRB-like events at higher frequencies. 
Though, more sophisticated studies of both the emission mechanism in three-dimensions and propagation effects in the binary wind will be required to clarify these points. 
The joint availability of all-sky-coverage at frequencies $\lesssim 20\, \rm GHz$, e.g., by SKA \cite{Weltman:2018zrl}, as well as early warning systems for neutron star coalescence \cite{James:2019xca,Sachdev:2020lfd,Wang:2020sda,Yu:2021vvm} might help to put this mechanism to a firm test.

\begin{acknowledgments}
The authors are grateful for discussions with K. Cleary,
G. Hallinan, M. Lyutikov, B. Metzger, K. Mooley, J. Nättilä, J. Stone and E. Quataert.
ERM gratefully acknowledges support from a joint fellowship
at the Princeton Center for Theoretical Science, the Princeton Gravity
Initiative and the Institute for Advanced Study. The simulations were
performed on the NSF Frontera supercomputer under grants AST20008 and
AST21006. AP acknowledges support by the National Science Foundation under grant No. AST-1909458. This work was performed in part at the Aspen Center for Physics, which is supported by National Science Foundation grant PHY-1607611, and was facilitated by Multimessenger Plasma Physics Center (MPPC), NSF Grant No. PHY-2206607. ERM also acknowledges the Extreme Science and Engineering Discovery Environment (XSEDE) \cite{Towns:2014qtb} through Expanse at SDSC and Bridges-2 at PSC through allocations PHY210053 and PHY210074. ERM further acknowledges the use of computational resources managed and supported by Princeton Research Computing, a consortium of groups including the Princeton Institute for Computational Science and Engineering (PICSciE) and the Office of Information Technology's High Performance Computing Center and Visualization Laboratory at Princeton University. ERM also acknowledges the use of high-performance computing at the Institute for Advanced Study.
This work has made extensive use of a number of software packages. Those
include AMReX \cite{amrex}, matplotlib \cite{Hunter:2007}, numpy
\cite{harris2020array} and scipy \cite{2020SciPy-NMeth}. 
\end{acknowledgments}

\newpage

\appendix

\textit{Appendix A: Methods} --
In this work, we simulate the global dynamics of the plasma by considering a perfectly
conducting electron-positron pair plasma \cite{Goldreich:1969sb} surrounding the two neutron stars
\cite{Hansen:2000am,Lyutikov:2018nti}.  Such global dynamics can be well captured using (resistive) force-free electrodynamics approaches (e.g., \cite{Palenzuela:2010nf,Parfrey:2013gza,Carrasco:2019aas}).
Building on our earlier work \cite{Most:2020ami,Most:2022ojl}, we model
the binary as two perfectly conducting spheres on a circular orbit. We
impose spin on the individual spheres by adjusting the local uniform
rotation rate. We exclude precessing motion and only include alignment of
the stellar and orbital rotation axes. Overall, we solve the equations of
general-relativistic electrodynamics
\cite{Baumgarte:2002vv,Palenzuela:2013hu} using a flat spacetime, which is
corotating at the orbital frequency \cite{Carrasco:2020sxg,Most:2022ojl}.
The effects of neglecting the gravitational potential (curved spacetime)
are negligible compared to the uncertainties in the emission modeling (see
emission section). We model the effects of a force-free pair plasma \cite{Goldreich:1969sb} using an effective current damping the electric field component parallel to the magnetic field \cite{Alic:2012df}, see also \cite{Spitkovsky:2006np}. This prescription for the current has been shown to introduce an effective resistive scale into the plasma \cite{Ripperda:2021pzt,Mahlmann:2021yws}, see also \cite{Li:2011zh}. The latter allows us to qualitatively capture reconnection in current sheets present in the simulations. Numerically, these equations are discretized using a fourth-order accurate finite-volume scheme \cite{mccorquodale2011high} with WENO-Z reconstruction \cite{Borges2008} and a Rusanov flux solver \cite{Rusanov1961a}. We further utilize the adaptive mesh-refinement infrastructure of the \texttt{AMReX} framework \cite{amrex}.  For the results presented in this work, we have considered two representative binary configurations, both at a separation distance of $52\, \rm km$. The first one is an irrotational system, with counter-oriented dipoles, and a ratio of the field strengths $B_1 \gg B_2$ at the surfaces of stars $1/2$, we choose $B_1 = 50 \times B_2$. The second system contains a counter-spinning neutron star with spin frequency $f=-300\, \rm Hz$ and a magnetic dipole that is inclined by $60^\circ$ with respect to the stellar spin axis.
The other star features a counter-aligned dipole with equal field
strengths, i.e. $B_1/B_2=1$. While most binary neutron star mergers
  are not expected to be rapidly spinning \cite{Bildsten:1992my}, we choose a spinning
  configuration to aid numerical simulation, since spin will affect both
  the emission period and in our case the location where the flare will
  interact with the orbital current sheet. The conclusions of our work
remain unaffected by this choice.
More details about the code and these types of setup can be found in
\cite{Most:2022ojl}.\\

\textit{Appendix B: Propagation effects} --
Fast-magnetosonic waves emitted during mergers of magnetic islands in the
reconnection layer propagate outwards on top of the magnetic pulse. These
waves can be released as radio emission when the pulse is decelerated at
the collision with the termination shock of the binary wind. However,
several processes can affect their propagation. Because the strength of the
magnetic field in the waves remains sub-dominant compared to the field in
the pulse, their conversion to strong shocks \cite{Beloborodov22b} and
intense scattering by magnetized particles \cite{Beloborodov22} remain
inefficient\footnote{Waves produced in the current sheet in the inner
magnetosphere will be strongly absorbed by these mechanisms.}. Here, we
apply the estimates by
\citet{Lyubarsky:2020qbp} performed in the context of reconnection-driven
radio bursts from magnetar magnetospheres, since this is the same mechanism
that we also invoke in the binary. 

In order to estimate the propagation effects, we first need to calculate
the properties of the binary wind and mass-loading of the pulse. The
particle flux in the wind is $\dot{N}=\mathcal{M}\mu c/(e R_{\rm LC}^2)$,
where $\mathcal{M}\sim 10^4$ is the plasma multiplicity\footnote{For
  $B_*\sim 10^{11} G$ and fast rotation of the binary, $P\sim {\rm 10}$ ms,
we assume a pulsar-like multiplicity in the magnetosphere, although this
point will require a separate investigation.} and $\mu$ is the magnetic
moment of stars. The magnetization parameter of the wind, the ratio of the
Poynting flux to the rest mass energy flux, is $\eta=B^2_{\rm LC}
R^2_{LC}/(\dot{N} m_e c)\sim (e B_{\rm LC} R_{\rm LC})/(\mathcal{M} m_e
c^2)\sim 5 \cdot 10^7 (B_{\rm LC}/(4\cdot 10^7{\rm G}))(P/4{\rm
ms})(10^4/M)$. Starting from the light cylinder, the wind linearly
accelerates with distance, until it reaches the magnetosonic surface at
$\sim R_{\rm LC} \eta^{1/3}\sim 7 \cdot 10^9 {\rm cm}$, beyond which the
acceleration is slow \citep{Beskin98}. The Lorentz factor of the wind at
this distance is $\gamma_{\rm w}=3\eta^{1/3}\sim 10^3$. It can further
accelerate because of the possible additional magnetic dissipation in the
stripes \cite[e.g.,][]{Lyubarsky01}. 

The pulse picks plasma particles from the inner magnetosphere, $N\sim
M\mu/(e R_{\rm LC})$, and drags them out. The corresponding magnetization
parameter of the pulse beyond the fast magnetosonic surface is
$\sigma_p=E_{p}/(Nm_e c^2 \gamma_w)\sim 3\cdot 10^7$, where $E_p\sim
10^{39}$ erg is the energy in the pulse. In the frame of the binary the
pulse propagates at the Lorentz factor $\gamma_p=2\Gamma \gamma_w\sim
6\cdot 10^3$ \citep{Lyubarsky:2020qbp}. 

The magnetization can be lower because of the additional pair production
during the interaction of the pulse and magnetospheric stripe close to the
light cylinder. The high value of the magnetic compactness parameter near
the reconnection layer at the light cylinder, $l_B=\sigma_T (B_{\rm flare}/\Gamma)^2 R_{LC}/(8\pi
m_e c^2)\sim 10^4$ \citep{Mahlmann:2022nnz}, suggests that enhanced pair
production is likely to occur in this regime, which would decrease the
pulse magnetization. Note that the value of the magnetic compactness and
efficiency of pair creation quickly drops with distance, as the field
strength in the flare drops and its Lorentz factor increases.This does
not affect the frequencies of emitted
fast-magnetosonic waves, because reconnection dynamics remains dominated by
synchrotron cooling, as long as $\gamma'_{\rm rad}=\sqrt{3e \beta_{\rm
rec}/(2 r^2_e B'_{\rm flare})}\sim 10^3 \ll \sigma_p$. The pulse also picks
up additional magnetic flux, $\phi \sim \xi B_{LC} R_{LC} R/(2\gamma^2_w)$,
where $\xi<1 $ is a dimensionless coefficient which describes the ratio of
the mean to the total field in the striped wind, and plasma from the wind.
It accumulates in a very thin shell, $\delta/l=\phi/\phi_{\rm flare}\sim
2\cdot 10^{-3}\xi (R/10^{12}{\rm cm})$, where $l=c\tau$ is the length of
the pulse, and does not affect the wave propagation in the body of the
pulse at the cyclotron absorption radius (see below). 

As the fast-magnetosonic waves propagate outwards, and the magnetic field
in the pulse drops, cyclotron absorption can become important. It happens
when the wave frequency in the pulse frame, $\omega'$ matches the
particle's gyrofrequency, $\omega'_B=eB'_{\rm pulse}/(m_e c)$. This happens
at a distance $R_c=e\sqrt{\mathcal L_{\rm EM}/c}/(2\pi\nu_{\rm FRB}m_e
c)=10^{12}{\rm cm} (\mathcal L_{\rm EM}/10^{42}{\rm erg/s})^{1/2}(15{\rm
GHz}/\nu_{\rm FRB})$. The optical depth can be calculated
as $\tau_c=\int 4\pi^2 (e^2/m_e c) \delta(\omega'-\omega'_B) N' dR'$, where
$N'$ is the pair density in the pulse co-moving frame. Using eq. (34) in
\citep{Lyubarsky:2020qbp}, one obtains $\tau_c \sim \pi^2 \nu_{\rm FRB}
R_c/(\sigma_p \gamma^2_p c)\sim (10^5/\sigma_p)(6\cdot 10^3/\gamma_p)^2$.
It means that cyclotron absorption is not important for $\sigma_p\gtrsim
10^5$.

Before reaching the cyclotron radius, the fast-magnetosonic waves may
steepen into shocks which can lead to their dissipation. In the pulse
frame, the steepening can occur at a distance $\sim \sigma_p (B'_{\rm
pulse}/\delta B')c/\omega'$. In the binary frame, it corresponds to a
distance $R_s\sim 4 \gamma^2_p \sigma_p c(\mathcal{L}_{\rm
EM}/\mathcal{L}_{\rm radio})^{1/2}/(2\pi \nu_{\rm FRB})\sim 4\cdot 10^9
\sigma_p {\rm cm}$, which is well outside the cyclotron radius for any
reasonable value of the pulse magnetization, $\sigma_p \gtrsim 10^2$. Thus,
the non-linear wave steepening is negligible. 

Another potentially important process is the non-linear decay of
fast-magnetosonic waves into a pair of Alfven waves or into an Alfven and
fast wave \citep{Lyubarsky:2020qbp}. In the inner wind zone, the
characteristic optical depth, $\tau_{\rm NL} \sim \int (\delta B'/B'_{\rm
pulse})^2 \omega' dr'/c$, is dominated by the contribution near the light
cylinder, when the wind is mildly relativistic. This estimate leads to
$\tau_{\rm NL}\sim (\mathcal{L}_{\rm EM}/\mathcal{L}_{\rm radio})(\omega
R_{LC})/(2c \gamma^2_p)\sim 50$, where $\gamma_p\sim 5$ near the light
cylinder. However, two important things may prevent efficient conversion
into Alfven waves. First, since they propagate along the magnetic field,
there are no pre-existing Alfven waves in the wind zone. In order for the
interaction to be efficient, Alfven waves need to grow from noise, which
requires substantial optical depth, $\sim 10$. Second, the above estimate
is an upper limit, since the non-linear interaction of fast and Alfven
waves decreases in the oblique limit, $k_{\perp}/k_{\parallel}>1$
\citep{TenBarge21}. To summarize, we expect only modest effect of
non-linear conversion processes on the wave propagation near the light
cylinder. Beyond the fast magnetosonic radius, the optical depth may
increase again, $\tau_{\rm NL}\sim (\mathcal{L}_{\rm
radio}/\mathcal{L}_{\rm EM})(\omega R)/(2c \gamma^2_p)\sim R/(2\cdot
10^{11}{\rm cm})$, which is nominally significant beyond the cyclotron
radius. However, at that distance the wave frequency exceeds the plasma
frequency,
$\omega'/\omega'_p=\sqrt{\sigma_p}\omega'/\omega'_B=\sqrt{\sigma_p}(R/R_c)\gg
1$, Alfven waves do not exist and the no non-linear interaction is absent.
Finally, beyond the cyclotron radius induced scattering can be important.
The estimate by \citep{Lyubarsky:2020qbp} shows $\tau_s \sim
(1/(\sigma_p\gamma^2_p))(\mathcal{L}_{\rm radio}/\mathcal{L}_{\rm
EM})(\omega_B/\omega)^4(\omega R/c)\sim 10 \sigma^{-1}_p (10^{12}{\rm
cm}/R)^3$, which is negligible for $\sigma_p \gtrsim 10$.

\newcommand{\apjl}{Astrophys. J. Lett.}
\newcommand{\apjs}{Astrophys. J., Supp.}
\newcommand{\mnras}{Mon. Not. Roy. Astron. Soc.}

\bibliography{inspire,non_inspire}%

\end{document}